\documentclass[twocolumn,twoside,slac_two]{revtex4}
\usepackage{graphicx}
\usepackage{fancyhdr}
\usepackage{hyperref}
\hypersetup{
    colorlinks=true,
    urlcolor=magenta,
}
\begin{document}

\title{Search for Neutrino Generated Air Shower Candidates with Energy More than 5$\cdot$10$^{18}$ eV and Zenith Angle $\theta$ $\geq$ 50$^\circ$}

%

\author{S. Knurenko, I. Petrov, A. Sabourov, Z. Petrov}
\affiliation{Yu. G. Shafer Institute of Cosmophysical Research and Aeronomy SB RAS, Yakutsk, Russia}

\begin{abstract}
Neutrino air showers can be formed in any part of the atmosphere passing a long was in the matter due to its physical properties. In general, air showers produced by neutrinos are highly inclined and formed near the ground level, i.e. young showers. Therefore, one should expect a large number of peaks in the signal of such air showers [1, 2, 3].

The goal of our work is to search for air shower candidates produced by neutrino. For this purposes, we analyzed large amount of data from scintillation detectors with different area and energy thresholds [1, 2].

Preliminary analysis of Yakutsk array data indicated the absence of air shower produced by neutrino, but it does not mean that such air showers does not exist. It’s going to need a further analysis of highly inclined showers. In order to do that improved methodology for recording and processing of air showers required.

\end{abstract}

\maketitle

\thispagestyle{fancy}


\section{Introduction}
\label{intro}
Air shower from neutrinos can be formed in any part of the atmosphere because of its physical properties, passing quite a long way in the matter. As a rule - it's strongly inclined showers formed near the detector, i.e. young showers. The basis of such shower is electron-photon component that is scattered at large angles and, therefore, has high latency particles relatively to particles formed in the shower core. Therefore, in such event should be expected a large number of peaks of the electrons in the signal scan from the scintillation detector [1, 2, 3].

The goal of our work was to search for extensive air showers (EAS) candidates produced by atmospheric neutrinos. In order to do this a large amount of air showers with a time sweep responses from scintillation detectors of different sizes and different energy thresholds has been analyzed [1, 2].

Preliminary results of data analysis of Yakutsk array indicates the absence of EAS events produced by neutrinos, but it does not mean that such EAS does not exist. It requires a further collection of strongly inclined showers and careful analysis. This requires improving the methodology for recording and processing of showers.

\section{Search Methodology of High Energy Gamma Rays and Astro-Neutrino}

\subsection{Longitudinal Development of Air Showers: Depth of Maximum Development X$_{max}$}
\label{sec-21}

\begin{figure}
\centering
\includegraphics[width=0.4\textwidth]{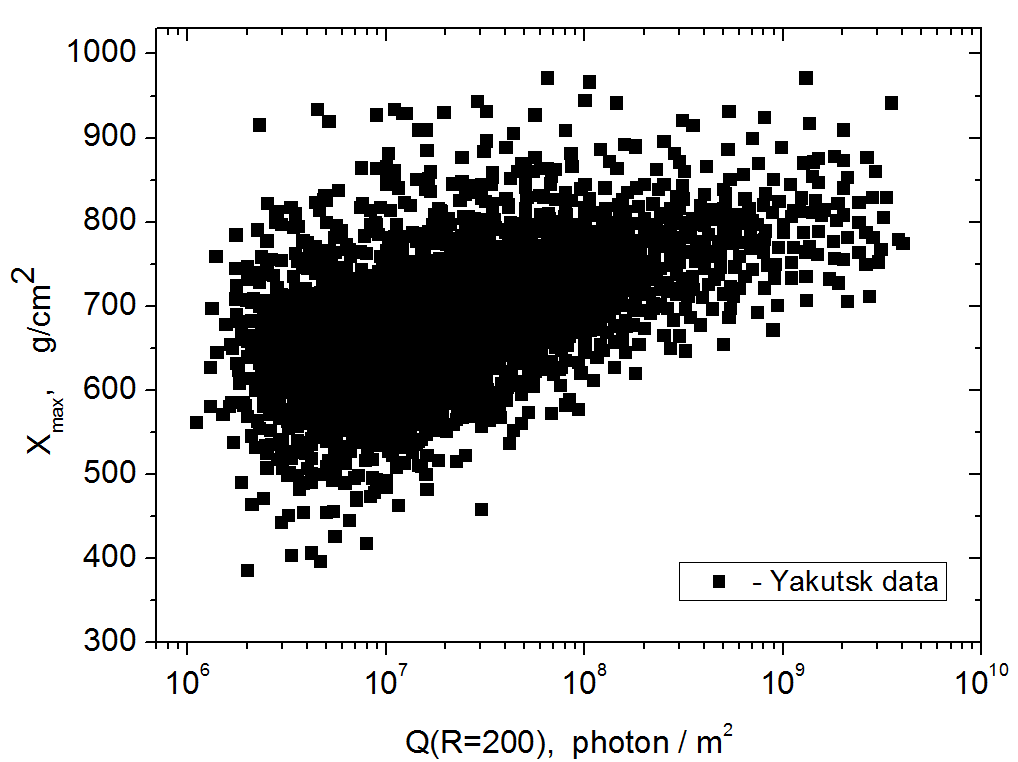}
\caption{Dependence of X$_{max}$ from classification parameter Q (200) - EAS Cherenkov light flux density at a distance of 200 m from the shower axis. Data obtained in 1973-2014}
\label{IP_fig1}       
\end{figure}

Longitudinal shower development at the Yakutsk reconstructed through registration of EAS Cherenkov light [4, 5], using the mathematical apparatus used when solving inverse problems [6, 7]. It allowed us to find connection parameters describing the cascade curve X$_{max}$, N$_{max}$, the width of the cascade q with directly measured air shower characteristics at the sea level, shower energy E, the total number of charged particles and the shape of the lateral distribution of the Cherenkov light R = lg (Qi / Qj), where Qi and Qj flux density of Cherenkov light at a selected distance from the shower axis [8]. With the use of this method, EAS database with reconstructed cascade curve was created. In addition, we selected air showers by X$_{max}$ to study depth of maximum offset dX$_{max}$/dlgE and its fluctuation $\sigma$$_{Xmax}$ in various energy intervals. Fig. \ref{IP_fig1} shows dependence of distribution of X$_{max}$ of individual showers from classification parameter of shower Q(200) - flux density of Cherenkov light at 200 m distance from shower axis.  Air shower energy can be determined by formula (\ref{IP_eq1}) obtained by energy balance method [9].

\begin{equation}
\label{IP_eq1}
E_0 = (1.78\pm0.44)\cdot10^{17}\cdot\left( \frac{Q(200)}{10^{7}}\right)^{1.01\pm0.04}
\end{equation}

\begin{figure}
\centering
\includegraphics[width=0.4\textwidth]{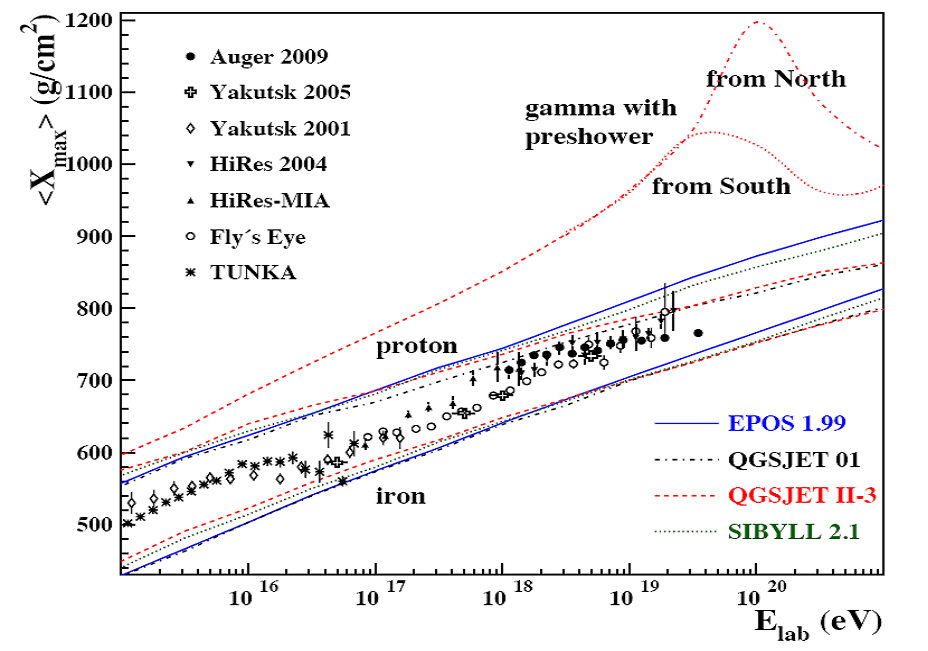}
\caption{Dependence of X$_{max}$ from energy. Experimental data comparison with calculations for different hadron interaction models (proton, iron nuclei and high-energy gamma ray).}
\label{IP_fig2}       
\end{figure}

Averaged Yakutsk data together with data from other arrays are shown in Fig. \ref{IP_fig2}. In addition, there is calculation for some hadron interactions models for primary nuclei (proton and iron) and high-energy gamma ray. Fig. \ref{IP_fig2} shows that height of shower produced by gamma ray is lower by 150-180 g/cm$^{2}$ than proton produced shower with energy 10$^{19}$ eV. In fact, cascade curve X$_{max}$ from gamma ray is near sea level at depth $\sim$950 g/cm$^{2}$. In this case, there is a narrow cascade mainly consisting of electrons and photons with a very low content of muons. This distinguishes shower produced by gamma ray from shower produced by proton or nucleus of any other element. We can assume that X$_{max}$ can be used as the first criterion to search for EAS produced by gamma ray.

\subsection{Amount of Muons in Air Shower}
\label{sec-22}

\begin{figure}[h]
\begin{minipage}[h]{0.8\linewidth}
\center{\includegraphics[width=0.9\linewidth]{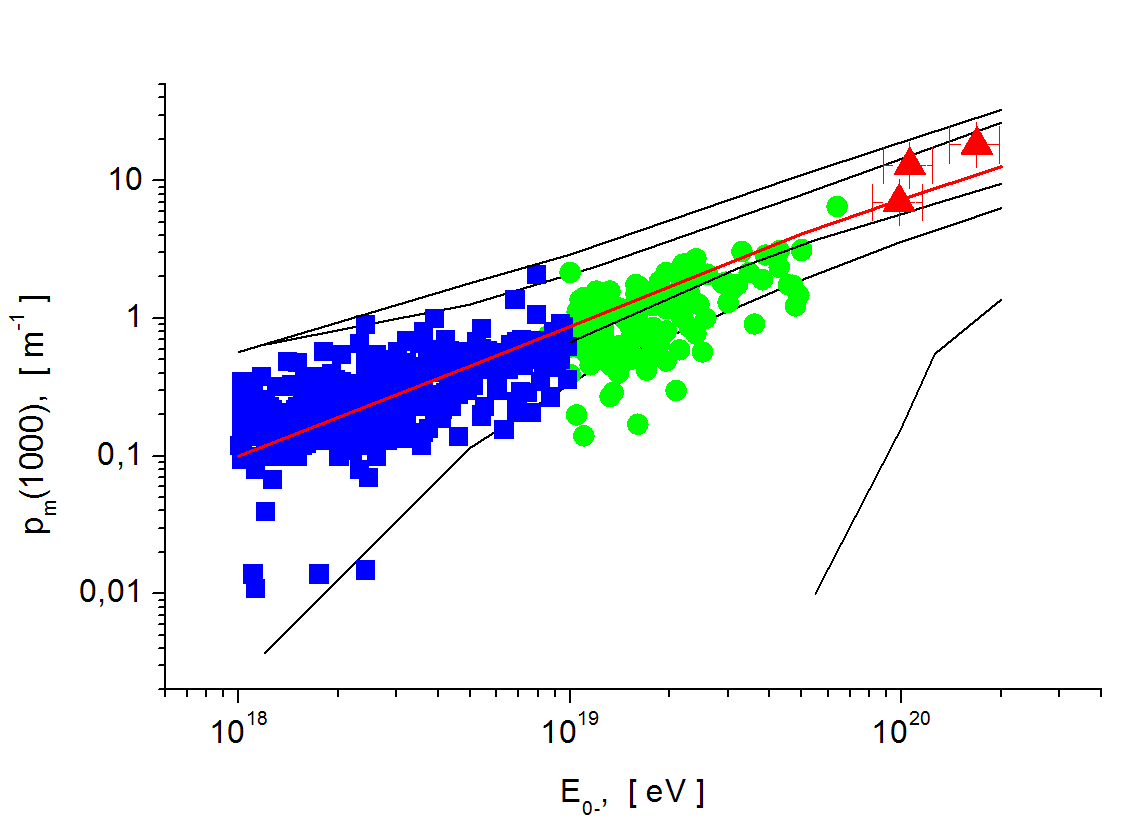} \\ a)}
\end{minipage}
\hfill
\begin{minipage}[h]{0.8\linewidth}
\center{\includegraphics[width=0.9\linewidth]{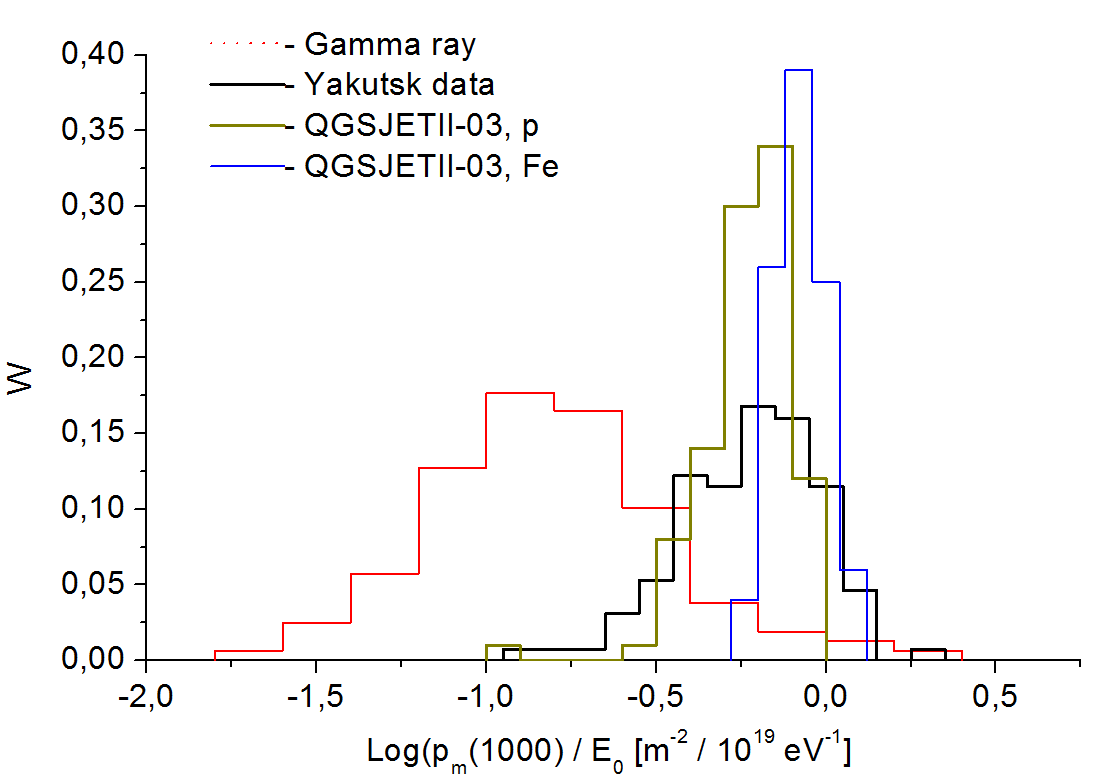} \\ b)}
\end{minipage}
\caption{ (a) Energy dependency of $\rho_{\mu}$(1000) for observed events with energy 10$^{18}$-10$^{19}$ eV (squares) and 10$^{19}$-10$^{20}$ eV (triangles). Expected $\pm$1$\sigma$ bounds of the distributions are indicated for proton, iron and gamma ray by different curves. (b) Fluctuations of $\rho_{\mu}$(1000)/10$^{19}$ value (eV) in showers with E$_{0}$ $>$ 10$^{19}$ eV compared to simulation results (QGSjetII-03 + UrQMD) for proton, iron and photon}
\label{IP_fig3}
\end{figure}

As mentioned in section \ref{sec-21} gamma ray produced shower consist small number of muons. If the array detects muons then from number of muons in the shower one can judge the nature of the primary particle i.e. its atomic weight (Fig. \ref{IP_fig3}a and \ref{IP_fig3}b).

In Yakutsk array the proportion of muons in the shower determined by relation of muon flux density at distances of 600 m and 1000 m to the total charged component $\rho_{\mu}$/$\rho_{\mu+e}$, since these parameters are measured with better accuracy than the total number of muons N$_{\mu}$ and charged particles N$_{\mu+e}$ in showers with total energy E$_{0}$ $\geq$ 10$^{18}$ eV. This will be the second criterion to search showers produced by neutral particles that includes gamma rays and neutrino.

For these purposes, there are muon detectors at the Yakutsk array with area of 1 m$^{2}$, 20 m$^{2}$ and 190 m$^{2}$, which records muons with a threshold energy $\varepsilon$$_{thr.}$ $\geq$ 1 GeV in vertical showers and $\varepsilon$$_{thr.}$ $\geq$ 2 GeV in inclined showers $\theta$ $\geq$ 60$^\circ$ [10].  The muon detectors are located on the array in such way that in each shower they are measuring distances in the range from 100 to 1500 m, which allows us to estimate the fraction of muons at a particular distance from the shower axis. In particular, this analysis uses the estimated proportion of muons in showers with energies $\geq$ 10$^{18}$ eV by ratio of normalized energy of the muon flux density to the density of charged particles measured by surface scintillation detectors. Experimental results are shown in Fig. \ref{IP_fig3}a, Fig. \ref{IP_fig3}b [11, 12, 13]. Lines are calculations performed by QGSJETII-03 model for proton, iron nucleus and the primary gamma ray. The calculations take into account the instrumental errors and for this reason, in Fig. \ref{IP_fig3}a double lines indicate the boundaries at 1$\sigma$ expected areas measurements of muons in the case of different mass composition of primary particles.

Fig. \ref{IP_fig3}a shows that there is overlapping region for gamma ray calculations (solid line) and the experimental data. In \ref{IP_fig3}b there is also similar overlap, which shows muon flux fluctuations at distance 1000 m from shower axis, normalized to the shower energy (dots). This suggests the existence of high-energy gamma rays in the flow of cosmic particles, which produces EAS in the Earth's atmosphere.

\subsection{Signal time sweep of surface and underground scintillation detectors}
\label{sec-23}

Air showers produced by different primary particles have a maximum development at different depth in the atmosphere. Because of this, part of secondary particles (mostly electrons) loses energy to ionization of the air and eliminated from the cascade process. Then on the sea level will arrive a certain type of particles: electrons, photons, muons in the case of inclined shower and only muons in the case of strongly inclined showers, which can be seen in the signal time sweep of scintillation detectors. For the primary gamma ray and neutrino, depth of maximum is going to be near the level of observation and we can expect scintillation detector response inherent in the electron-photon component of the shower. This is another criterion by which we can select showers produced by neutral particles.

\begin{figure}[h]
\begin{minipage}[h]{0.8\linewidth}
\center{\includegraphics[width=0.78\linewidth]{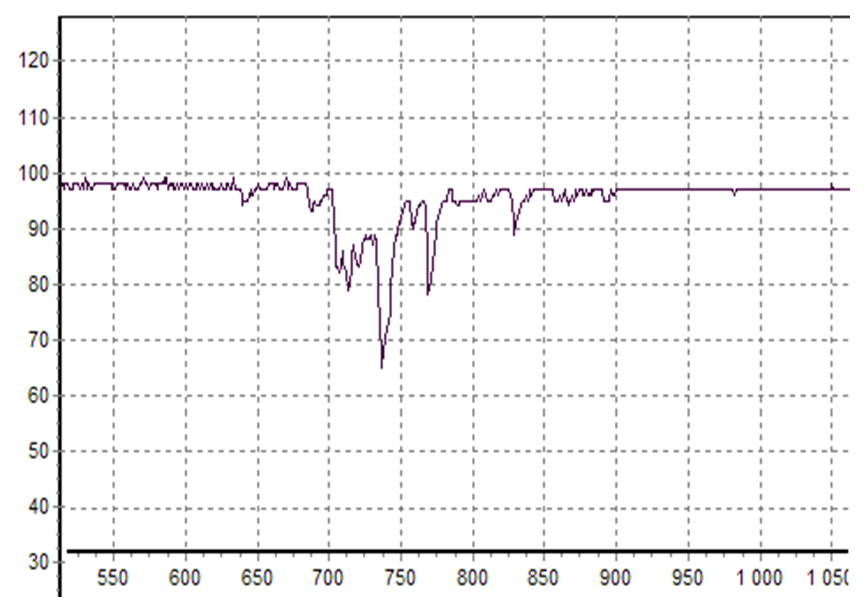} \\ a)}
\end{minipage}
\hfill
\begin{minipage}[h]{0.8\linewidth}
\center{\includegraphics[width=0.78\linewidth]{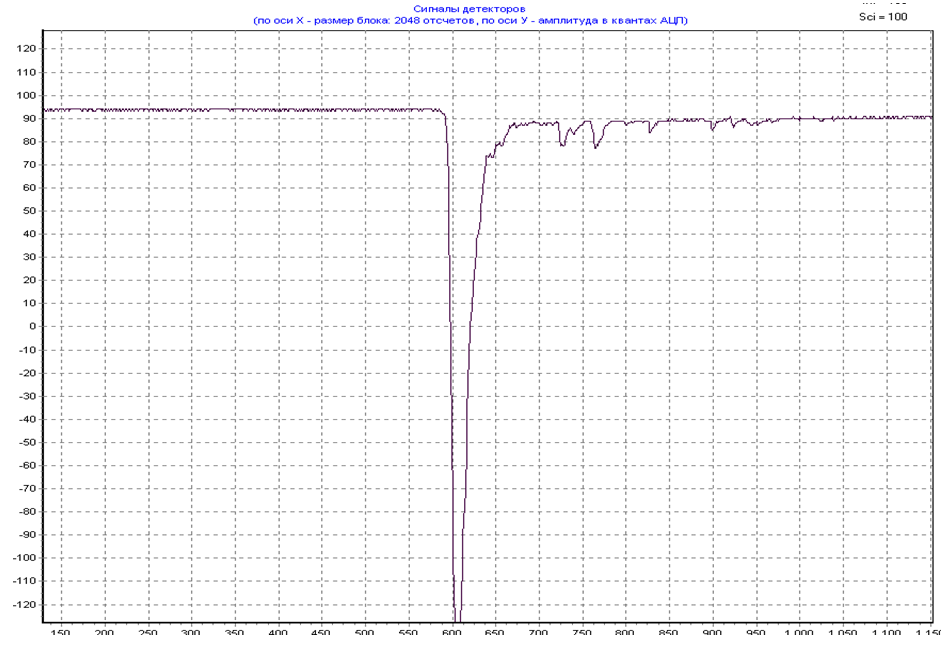} \\ b)}
\end{minipage}
\caption{ (a) Vertical air shower pulse. E$_{0}$ = 1.7$\cdot$10$^{19}$ eV, $\theta$ = 18$^\circ$, R =1298 m. Detector with area s = 2 m$^{2}$ and threshold $\varepsilon$$_{thr}$ $\geq$ 10 MeV. (b) Inclined air shower event. E$_{0}$ = 2.4$\cdot$10$^{19}$ eV, $\theta$ = 56$^\circ$, $\psi$ = 200$^\circ$, R =1000 m. }
\label{IP_fig4}
\end{figure}

Examples of scintillation detector responses in the case of vertical and inclined showers are given in Fig. \ref{IP_fig4}a and Fig. \ref{IP_fig4}b. Fig. \ref{IP_fig4}a and Fig. \ref{IP_fig4}b shows that the shape of the pulse in each case is different. Vertical showers have many peaks, where each peak corresponds to a group or a single particle. As can be seen, all of the particles are distributed in time, i.e. they arrive with different delays with respect to the first particle. Most likely, it is the electrons scattered in the shower subcascades.

From Fig. \ref{IP_fig4}b it is seen that in strongly inclined showers the pulse structure is different from vertical showers. It is clear single and narrow pulse. The compactness of the arrival of these particles indicates that these particles are produced in the first interactions of primary particle with air nuclei and in the course of decay processes of $\pi\pm$ - mesons, i.e. they are muons.

It must be noted that the signal shape asymmetry significantly expressed in showers with zenith angles $\theta$ = 40$^\circ$ - 55$^\circ$. For large zenith angles, only narrow single pulse is observed in the signal scan from air shower.

\section{Experimental Data Analysis}

\begin{table*}[htb!]
\centering
\caption{Strongly inclined showers with energy E $\geq$ 1$\cdot$10$^{19}$ eV and $\theta \geq$ 60 $^{\circ}$}
\label{tab-2}

\begin{tabular}{|l|l|l|l|l|l|l|l|}
\hline
Date          & $\cos\theta$ & $\theta$ & lg E  & Station number & number of peaks & $\rho_{\mu}/\rho_{\mu+e}$ & QGSjetII-03 \\\hline
11.05.2000    & 0.266        & 74.6     & 19.63 & 33             & 1               & 0.91$\pm$0.09             & 0.84$\pm$0.03/0.89$\pm$0.03/0.20$\pm$0.03\\\hline
11.01.2002    & 0.322        & 71.2     & 19.08 & 27             & 1               & 0.92$\pm$0.10             & 0.84$\pm$0.03/0.89$\pm$0.03/0.20$\pm$0.03\\\hline
11.02.2002    & 0.330        & 70.7     & 19.04 & 27             & 1               & 0.94$\pm$0.11             & 0.84$\pm$0.03/0.89$\pm$0.03/0.20$\pm$0.03\\\hline
14.04.2002    & 0.319        & 71.4     & 19.10 & 32             & 1               & 1.11$\pm$0.16             & 0.84$\pm$0.03/0.89$\pm$0.03/0.20$\pm$0.03\\\hline
11.05.2006    & 0.437        & 64.1     & 19.05 & 42             & 1               & 1.06$\pm$0.09             & 0.84$\pm$0.03/0.89$\pm$0.03/0.20$\pm$0.03\\\hline
12.10.2006    & 0.445        & 63.6     & 19.17 & 49             & 1               & 1.17$\pm$0.12             & 0.84$\pm$0.03/0.89$\pm$0.03/0.20$\pm$0.03\\\hline
03.01.2007    & 0.495        & 60.3     & 19.43 & 35             & 1               &                           & 0.84$\pm$0.03/0.89$\pm$0.03/0.20$\pm$0.03\\\hline
15.05.2007    & 0.498        & 60.1     & 19.29 & 26             & 1               &                           & 0.84$\pm$0.03/0.89$\pm$0.03/0.20$\pm$0.03\\\hline
10.03.2008    & 0.425        & 64.8     & 19.18 & 49             & 1               &                           & 0.84$\pm$0.03/0.89$\pm$0.03/0.20$\pm$0.03\\\hline
21.02.2009    & 0.257        & 75.1     & 19.31 & 18             & 1               & 1.04$\pm$0.07             & 0.84$\pm$0.03/0.89$\pm$0.03/0.20$\pm$0.03\\\hline
21.03.2009    & 0.280        & 73.7     & 19.11 & 28             & 1               &                           & 0.84$\pm$0.03/0.89$\pm$0.03/0.20$\pm$0.03\\\hline
10.04.2009    & 0.401        & 66.4     & 19.17 & 41             & 1               &                           & 0.84$\pm$0.03/0.89$\pm$0.03/0.20$\pm$0.03\\\hline
25.05.2009    & 0.298        & 72.7     & 19.24 & 20             & 1               &                           & 0.84$\pm$0.03/0.89$\pm$0.03/0.20$\pm$0.03\\\hline
24.12.2009    & 0.402        & 66.3     & 19.21 & 44             & 1               &                           & 0.84$\pm$0.03/0.89$\pm$0.03/0.20$\pm$0.03\\\hline

\end{tabular}
\end{table*}
In this work, for the pulse shape analysis we selected EAS events with $\theta\geq$60$^\circ$ and energy above 5$\cdot$10$^{18}$ eV. It was necessary to shower axis to be within a circle with a radius of 1000 meters from the center of the Yakutsk array, where much of the monitoring stations and all underground stations for muon detection are located. In this case, the accuracy of the shower axis determination ($\sigma$$_x$ = 25 m, $\sigma$$_y$ = 35 m) was about the same and the best for all selected events. This also applies to the determination of the basic parameters of showers, such as the energy flux density of charged particles and muons at distances of 600 m and 1000 m, the zenith ($\sigma$$_\theta$ = 1.5$^\circ$) and azimuth ($\sigma$$_\varphi$ = 3.5$^\circ$) angles. Selection of showers included not only the requirement to pulse shape, but also its amplitude. Therefore, in the analysis we included showers with the pulse amplitude 5$\sigma$ more than the average noise signal on the scan. Share of muons was determined by the ratio of the density of muons registered at distances of 600 m and 1000 m from the shower axis with respect to the flow of charged particles, registered at the same distances. The accuracy of determination of each component in the range of (15 - 25) $\%$ [14, 15].

Selected according to the criteria showers formed the basis of the data for analysis. A separate list made up showers with energies above 1$\cdot$10$^{19}$ eV. The sample made, since 2000 to 2014 and included the showers registered with both surface and underground detectors (charged particles and muons).

\begin{figure}[h]
\begin{minipage}[h]{0.8\linewidth}
\center{\includegraphics[width=0.78\linewidth]{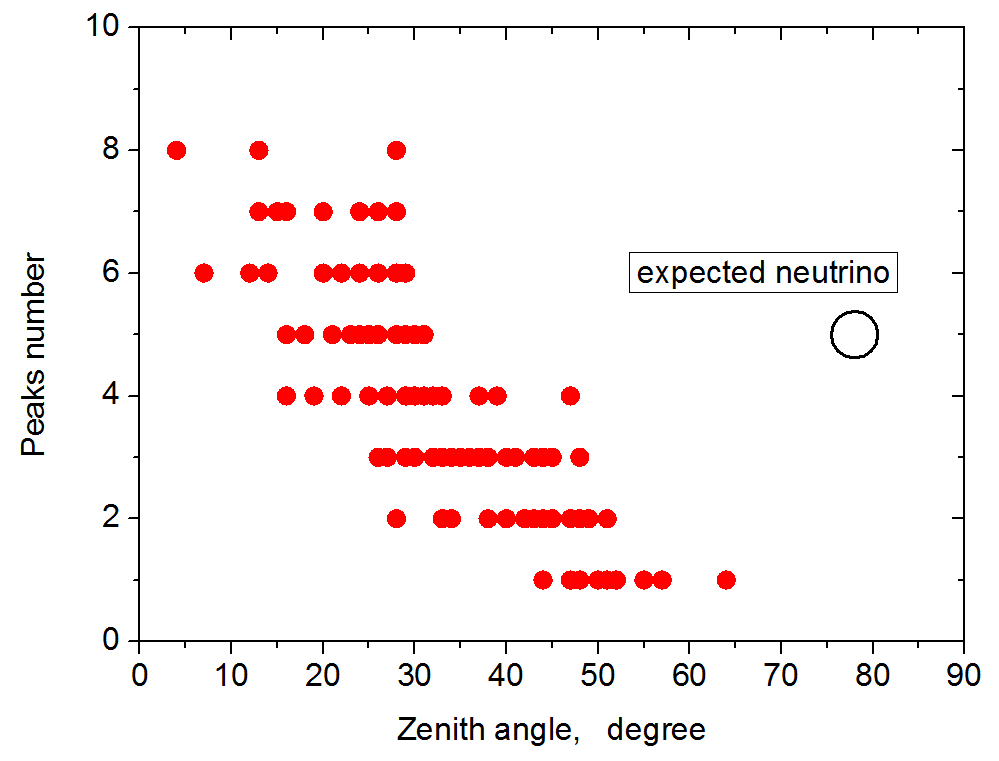} \\ a)}
\end{minipage}
\hfill
\begin{minipage}[h]{0.8\linewidth}
\center{\includegraphics[width=0.78\linewidth]{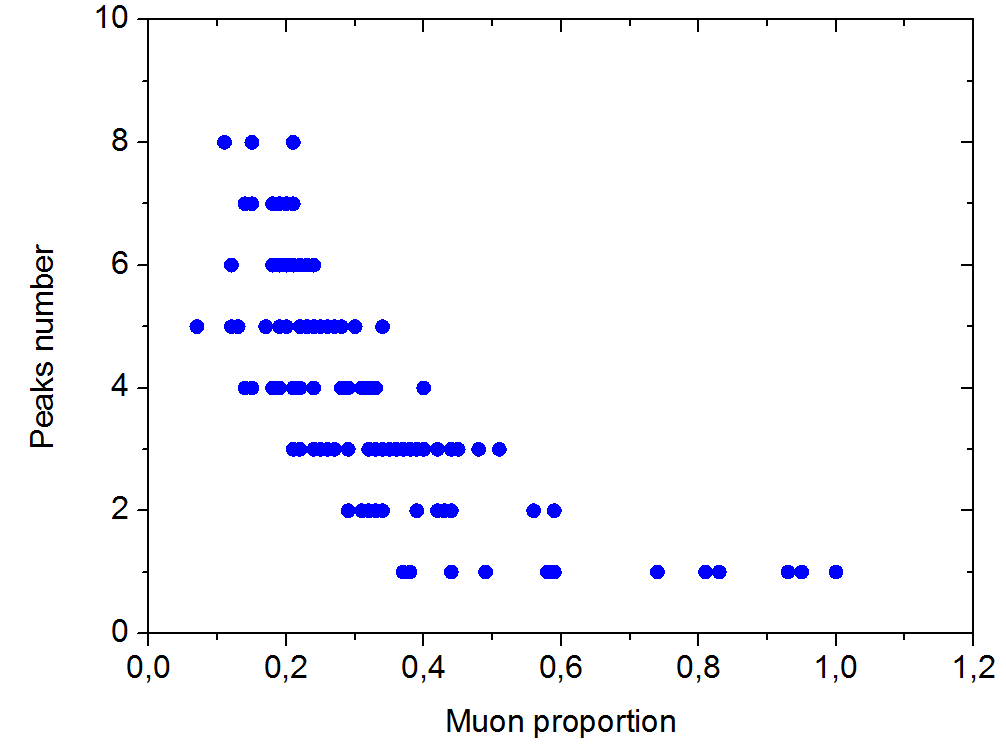} \\ b)}
\end{minipage}
\caption{ Number of peaks in time sweep of scintillation detector. a) Surface scintillation detector with threshold 10 MeV plotted against zenith angle b) underground scintillation detector with $\varepsilon_{thr.}$ $\geq$ 1 GeV. }
\label{IP_fig8}
\end{figure}

Characteristics of the largest showers are given in Table 1. These and other showers were included in this analysis to search for EAS candidates produced by neutral particles.

In Table 1, $\rho_{\mu}/\rho_{\mu+e}$ for the distance 600 m from shower axis. In the case of QGSjetII-03, calculations for p/Fe/$\gamma$ for energy E=10$^{19}$ eV and $\theta$ $\geq$ 60$^\circ$.

Fig. \ref{IP_fig8} shows dependence of number of peaks in the time sweep response of the scintillation detector from the zenith angle of the shower arrival (Fig. \ref{IP_fig8}a) and from the proportion of muons at a distance of 600 meters from the EAS axis (Fig. \ref{IP_fig8}b).

It can be seen that the number of peaks in signal scan (Fig. \ref{IP_fig8}a) strongly depends on the zenith angle. In the case of vertical showers with $\theta$ $\leq$ 30$^{\circ}$ and moderate distances from the shower axis, the number of peaks ranges from eight to four, and starting from zenith angles $\theta$ $\geq$ 59 $^\circ$ there is only single peak in signal scan. Proportion of muons increases significantly with zenith angle and reaches almost 100$\%$ for $\theta$ $\geq$ 60 $^\circ$. The observed pattern can be used as a sort of search criteria for air shower produced by neutral particle. For example if EAS has depth of maximum near sea level, X$_{max}$ = (800-1000) g/cm$^{2}$, amount of muons is small and there are large amount of peaks in signal scan at zenith angle $\theta$ $\geq$ 60 $^\circ$ we can assume that this shower with a high probability produced by neutral particle - gamma ray or neutrino. In Fig. \ref{IP_fig8}a, this region is marked by large circle. Thus, suggested in this work search criteria for primary particles with characteristics different from protons and other nuclei may be suitable for such analysis.

\section{Conclusion}

Over 15 years of continuous observations at the Yakutsk 1944 EAS events with energy above 5$\cdot$10$^{18}$ eV and $\theta$ $\geq$ 60 $^\circ$ were registered. With methodology suggested above, comprehensive analysis of charged particles, muons and Cherenkov light data was carried out including time sweep in underground and surface scintillation detectors. Dependence of shape scintillation detectors response in time sweep from the zenith angle. A characteristic feature was that where were many peaks in vertical showers and only single peak in strongly inclined showers [2]. Use of Cherenkov component of the shower in the analysis and reconstruction of shower longitudinal development [5, 8] indicated direct relationship of the peaks in time sweep of scintillation detector with longitudinal development of the shower and amount of muons in the shower. All this is well illustrated by the data shown in Fig. \ref{IP_fig8}a, Fig. \ref{IP_fig8}b.

Multi component analysis of air showers with use of above described criteria, found no showers produced by gamma ray or neutrino. At the same time, QGSJETII-03 model calculations for the primary protons and iron nuclei and gamma ray (Fig. \ref{IP_fig2}, Fig. \ref{IP_fig3}a and Fig. \ref{IP_fig3}b) tells that if fluctuations of the muon measurements within 1$\sigma$ is taken into account, the probability of detecting air shower produced by neutral particles exist. "Muonless" showers detected at the Yakutsk array can be considered as candidates for such showers [16].

%
%
%

\end{document}